\documentclass[doublecol]{epl2}

\title{On the scaling behaviour of cross-tie domain wall structures in patterned NiFe elements}
\shorttitle{Scaling behaviour of cross-tie domain wall structures} 

\author{N. Wiese\inst{1} \and S. McVitie\inst{1} \and J.N. Chapman\inst{1} \and A. Capella-Kort\inst{2} \and F. Otto\inst{2}}

\shortauthor{N. Wiese \etal}

\institute{
  \inst{1}University of Glasgow, Department of Physics and Astronomy, Glasgow G12 8QQ, United Kingdom \\
  \inst{2}Universit{\"a}t Bonn, Institute for Applied Mathematics, Wegelerstr. 10, 53115 Bonn, Germany
}


\abstract{
The cross-tie domain wall structure in micrometre and sub-micrometre wide patterned elements of NiFe, and a thickness range of 30 to 70nm, has been studied by Lorentz microscopy. Whilst the basic geometry of the cross-tie repeat units remains unchanged, their density increases when the cross-tie length is constrained to be smaller than the value associated with a continuous film. This occurs when element widths are sufficiently narrow or when the wall is forced to move close to an edge under the action of an applied field. To a very good approximation the cross-tie density scales with the inverse of the distance between the main wall and the element edge. The experiments show that in confined structures, the wall constantly modifies its form and that the need to generate, and subsequently annihilate, extra vortex/anti-vortex pairs constitutes an additional source of hysteresis.
}

\begin{document}
\maketitle

\section{Introduction}
\label{intro}
Thin films of soft magnetic materials have been the subject of many investigations for almost 50 years. Such films, when patterned into small elements, are used extensively in magnetic devices, particularly magnetic sensors, transducers and magnetic random access memories.\cite{Berg99,Gallagher97} Domain walls with a variety of different structures have been found in these films, the wall type depending on material parameters, such as thickness, magnetisation, exchange constant and anisotropy. 

\begin{figure}
\begin{center}
\includegraphics[width=8.6cm]{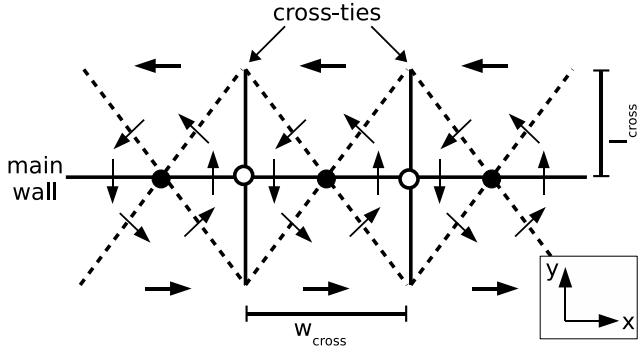}
\caption{\label{figure0}Schematic of a cross-tie domain wall.}
\end{center}
\end{figure}

The cross-tie domain wall (XDW) is one wall type encountered frequently. It was first observed in continuous films of soft magnetic materials, and has been studied by various experimental techniques, such as the Bitter technique \cite{Huber58}, electron microscopy \cite{Schueppel62} and Kerr microscopy.\cite{Hubert98} A XDW is shown schematically in figure \ref{figure0} and consists of a main DW, separating two antiparallel magnetic domains. The structure of the main wall varies continuously along its length, comprising alternating N\'{e}el and Bloch sections. XDWs are found in films with low anisotropy in an intermediate thickness range. Below the lower bound, N\'{e}el walls exist and above the upper bound, the asymmetric Bloch wall is found \cite{Hubert98}. Close examination of fig. \ref{figure0} shows that adjacent N\'{e}el sections are of opposite chirality leading to a succession of local vortices separated by low angle N\'{e}el walls, the so-called cross-ties. At the centre of each vortex and where the main wall and the cross-ties intersect, the magnetisation is directed perpendicular to the plane of the film so that, locally, the sense of moment rotation across the wall is Bloch-like. In continuous permalloy (Ni$_{80}$Fe$_{20}$) films, the thickness range over which the XDW is found is approximately 30 to 90 nm \cite{Middelhoek63}. Whilst this thickness range is considerably greater than that frequently employed in sensors, it is appropriate for magnetic elements in such transducers as magnetic recording write heads and is rapidly becoming the preferred range for the soft underlayer in perpendicular recording media \cite{Shi06}.

An important feature for theoretical models as well as for experiments is the multi-scale characteristics of the XDW. Cross-ties in continous films extend typically over several micrometres, whereas the core of the N\'{e}el segments is typically in the range of the exchange length, of approximately 5 nm in NiFe.\cite{Hubert98} Therefore, predicting the behaviour of XDWs is a challenge for micromagnetic as well as for analytical models, and reliable experimental data are needed to test any models developed. Most experimental studies of XDWs to date have been obtained from unpatterned films \cite{Ploessl93}. However, as interest increasingly turns to the behaviour of DWs in elements, it is here that our attention is focused. Of particular concern is what happens to the XDW when the element width is smaller than the cross-tie length in a continuous film. To study this we have used Lorentz microscopy, a technique well suited to determine the magnetisation distribution in micrometre and sub-micrometre sized elements \cite{Liu04}.

\begin{figure}
\begin{center}
\includegraphics[width=8.6cm]{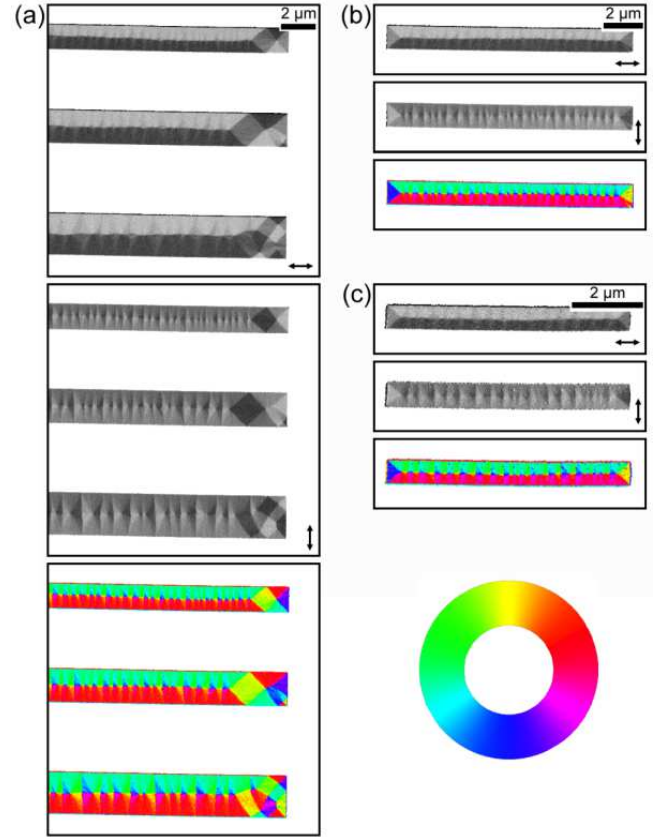}
\caption{\label{figure1}[colour online] DPC images of elements with constant aspect ratio, $\ell / w = 10$, and varying width of (a) $1.5 < w < 2.5\mu$m, (b) $w = 1.25\mu$m, and (c) $w = 0.5\mu$m. The direction of sensitivity has been chosen parallel and perpendicular to the long axis of the elements, respectively, as indicated by the arrows. The third image in each set shows the angular distribution of the induction within the elements, calculated from the two vector components.}
\end{center}
\end{figure}

\section{Experiment}
To investigate the variation of the XDW structure over a wide size range, elongated rectangular NiFe elements of different thickness and with in-plane dimensions (width $w$, length $\ell$) between $100$nm and 10$\mu$m, were fabricated using electron beam lithography. Two PMMA layers of different molecular weight were spun on electron transparent TEM substrates, after which e-beam exposures were carried out at 100kV using a vector scan lithography tool (Leica EBPG5-HR100). Following development of the exposed resist, NiFe was thermally evaporated after which the metal film on top of the unexposed resist was lifted off in warm acetone, leaving the required array of elements. Three samples with thicknesses 30, 50 and 70 nm were prepared. In addition to the patterned films, unpatterned films were deposited simultaneously. Hysteresis loops from these, obtained using the magneto-optic Kerr effect, showed that the films exhibit a low anisotropy field of $H_{\mbox{\scriptsize k}} \approx 7$Oe, and a low coercive field of $H_{\mbox{\scriptsize c}} < 1$Oe.

Lorentz microscopy was used to image the magnetic domain structures with the elements in their remanent states and when subjected to an in-situ applied field. The experiments were carried out in a modified Philips CM20 FEG electron microscope, equipped with a special set of Lorentz lenses, a digital CCD camera and differential phase contrast (DPC) detector. Both the Fresnel and DPC imaging modes were used. The set-up and the Lorentz imaging modes are described in detail elsewhere \cite{Chapman84}.

\section{Results and Discussion}
A XDW, running parallel to the long axis for most of the element length, could be generated reproducibly in the majority of elements with $t\geq 30$nm, using a strong magnetic field, $H_y$, along the short axis of the rectangular element. For elements with $w \leq 200$nm no XDW was formed, the remanent state being a S-like domain state \cite{Liu04}. Figure \ref{figure1} shows DPC image pairs, sensitive to induction components parallel and perpendicular to the element axes, obtained after application of $H_y=4$kOe. For these elements, $t=50$nm, $w$ was in the range of $0.5\mu \mbox{m} < w < 2.5\mu \mbox{m}$, and there was a constant aspect ratio of $\ell / w = 10$. Examination of the images shows that the geometry of the XDW was largely unaltered by the reduction in element width but the dimensions of the XDW changed significantly as the width was reduced. Indeed, $w_{\mbox{\scriptsize cross}}$ decreased from a value of $1.6\mu m$ for $w=10\mu$m to a value of $155$nm in elements with $w=0.4\mu$m. It is worth noting that the former value for wcross is close to the value found in unpatterned films with t = 50 nm.

\begin{figure}
\begin{center}
\includegraphics[width=8cm]{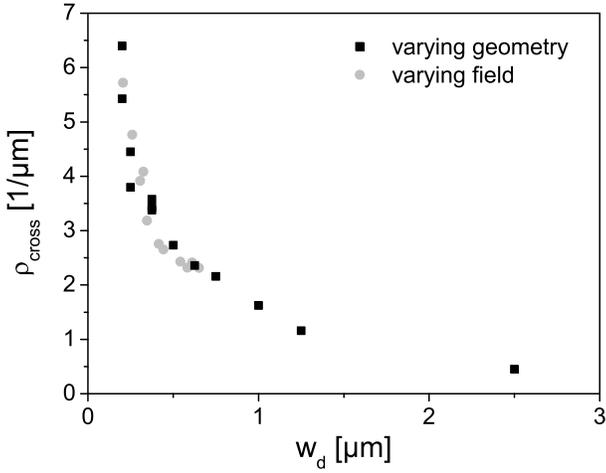}
\caption{\label{figure2}Cross-tie density, $\varrho_{\mbox{\scriptsize cross}}$, vs. domain width, $w_{\mbox{\scriptsize d}}$, determined for elements with $t=50$nm. Data with varying dimensions are extracted from elements with constant aspect ratio, $\ell / w = 10$, and $0.4<w<5\mu$m. If a magnetic field, $H_x$, is applied along the long axis of the element, the same dependence of $\varrho_{\mbox{\scriptsize cross}}$ vs. $w_{\mbox{\scriptsize d}}$ is obtained, as shown for a rectangular element with $w=1.25\mu$m and $\ell=10\mu$m.}
\end{center}
\end{figure}

From fig. \ref{figure1} and other similar images, the average cross-tie density, $\varrho_{\mbox{\scriptsize cross}}=1/w_{\mbox{\scriptsize cross}}$, was evaluated and this is plotted in fig. \ref{figure2} as a function of the distance of the XDW from the element edge, $w_{\mbox{\scriptsize d}}$. As the XDW runs down the centre of the element in the remanent state, $w_{\mbox{\scriptsize d}} \approx w/2$.

\begin{figure}
\begin{center}
\includegraphics[width=8.6cm]{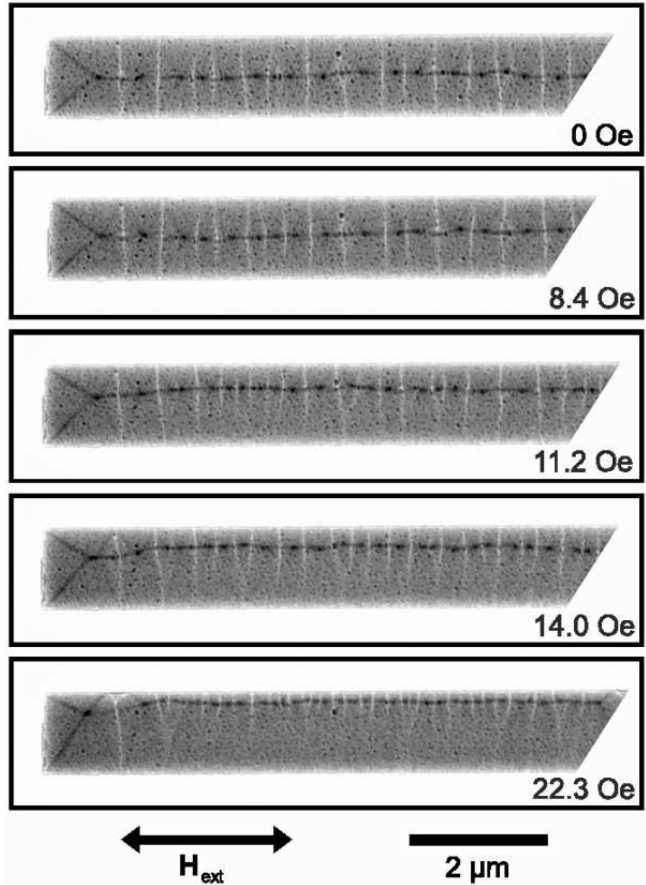}
\caption{\label{figure3}Sequence of Fresnel images obtained during an in-situ magnetising experiment of the sample with $t=50nm$. External magnetic fields, $H_{\mbox{\scriptsize ext}}$, have been applied parallel to the long axis of the rectangular elements, as indicated.}
\end{center}
\end{figure}

In the next set of experiments, low fields, $H_x$, were applied parallel to the long axis of the elements to move the XDW towards one of the element edges. Figure \ref{figure3} shows a sequence of Fresnel images for an element of $1.25 \times 10 \mu$m$^2$ and a thickness of $t=50$nm over the field range $0$Oe$<H_x<22.3$Oe. The width, $w_{\mbox{\scriptsize d}}$, of the domain with magnetisation antiparallel to $H_x$ decreased with increasing $H_x$, while new vortex/anti-vortex pairs were created, thereby increasing $\varrho_{\mbox{\scriptsize cross}}$. In fig. \ref{figure2}, the evaluated data for the cross-tie density obtained from the in-situ magnetising experiments is presented. The results show an identical dependence of the cross-tie density to that displayed by elements of decreasing width in their remanent state, discussed previously. Thus the key parameter is confirmed to be the distance of the XDW from the element edge. 

It should be emphasised that in the case of the data extracted in the field varying experiments, $w_{\mbox{\scriptsize d}}$ is the distance between the XDW and the element edge nearest to it. Its distance from the further edge is unimportant as was verified by carrying out experiments on elements of different width. In the case of the remanent state data, there is no potential for confusion in that the wall is centrally sited and hence equidistant from either edge. The reason for there being the single key parameter identified above will become clear in the analysis that follows shortly.

\begin{figure}
\begin{center}
\includegraphics[width=8cm]{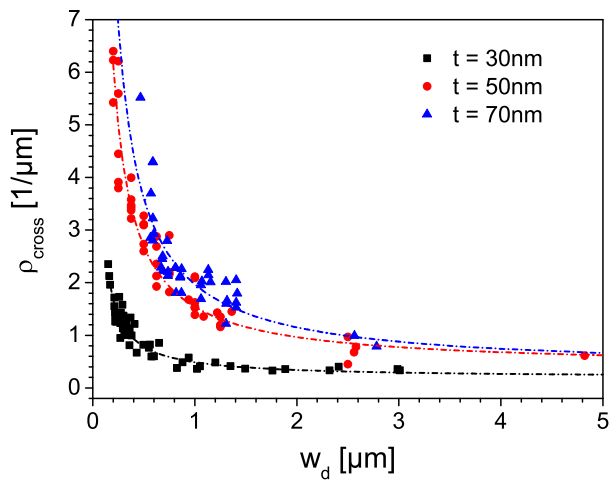}
\caption{\label{figure4}[colour online] Cross-tie density, $\varrho$, plotted against the domain width, $w_{\mbox{\scriptsize d}}$, for elements with thickness $t=30, 50,$ and $70nm$. The lines are least-squares fits as described in the text.}
\end{center}
\end{figure}

In fig. \ref{figure4}, $\varrho_{\mbox{\scriptsize cross}}$ is shown as a function of $w_{\mbox{\scriptsize d}}$ for elements of thickness $t=30, 50$ and $70$nm. The data were obtained using both methods described previously; i.e. by varying the element width and taking measurements in the remanent state, and by reducing the separation of the XDW and element edge by an external magnetic field. All data in fig. 3 is subsumed in fig. 5. The dependence of $\varrho_{\mbox{\scriptsize cross}}$ on $w_{\mbox{\scriptsize d}}$ is similar for all three thicknesses, whilst for a fixed value of $w_{\mbox{\scriptsize d}}$, $\varrho_{\mbox{\scriptsize cross}}$ increases with element thickness.

For comparison with the experimental results, we looked into numerical methods and existing analytical models. The former experience difficulties due to the need to discretise on the scale of the exchange length (or smaller) of the material leading to excessive array sizes in comparison with the computational power that is available today. Of the analytical theories, the earliest, due to Middelhoek \cite{Middelhoek63}, has been shown to be in good agreement with experiment for continuous films. However, it is inapplicable to the present case where the lateral dimensions of the elements are confined to sizes comparable to or smaller than the typical dimensions of the cross-ties in unpatterned films. Recent analytical models, though, do provide a more attractive route in that they allow comments to be made on the scaling behaviour of XDWs \cite{Alouges02,DeSimone04}. Here, Alouges et al. \cite{Alouges02} take as an ansatz constancy in the shape of the basic repeat unit in a cross-tie wall and DeSimone et al. evaluate the optimum cross-tie spacing \cite{DeSimone04}. In this approach the ratio of $\ell_{\mbox{\scriptsize cross}}$ to $w_{\mbox{\scriptsize cross}}$, fig. \ref{figure0}, is constant. Hence if $\ell_{\mbox{\scriptsize cross}}$ is forced to reduce due to constraints imposed by $w_{\mbox{\scriptsize d}}$, it follows that $w_{\mbox{\scriptsize cross}}$ and so $\varrho_{\mbox{\scriptsize cross}}$ will change accordingly to preserve the favoured geometry. Thus under constraints imposed either by element width or the proximity of an edge, the cross-tie density should vary as
\begin{equation}
     \varrho = 1/w_{\mbox{\scriptsize cross,0}} + B/w_{\mbox{\scriptsize d}}
\label{fit}
\end{equation}

Here $w_{\mbox{\scriptsize cross,0}}$ is the asymptotic value of the cross-tie separation, namely the value found in continuous films and $B$ depends on the geometry of the pattern. The dashed lines in fig. \ref{figure4} show the resulting least-squares fits to the data. For elements with $t=30$nm and $t=50$nm, the fits are good, the goodness of fit parameters $R^2$ being 0.90 and 0.92, respectively. For the $t=70$nm case, the fit is less good, $R^2$ assuming a value of 0.76. Values of $w_{\mbox{\scriptsize cross,0}}$ for the three cases in ascending order of film thickness were $5.0\mu$m, $2.6\mu$m and $3.0\mu$m. The values for the $t=50$ and $70$nm cases are in good accord with experimental data taken from $60$nm thick continuous NiFe films, where a value of $2.8\mu$m has been reported \cite{Ploessl93}. Although the fit to the data for $t=70$nm is much poorer, the general trend whereby the families of curves in fig. \ref{figure4} sit above each other according to their respective thicknesses is consistent with experimental results reported in \cite{Hubert98}.


\section{Conclusion}
Our experiments have shown that the structure of a XDW is not fixed by material parameters but changes when constraints are imposed on the length of the cross-ties. The constraint may be imposed by the width of a patterned element or may be the result of an applied field driving a XDW close to the film or element edge. Indeed, the latter result emphasises that the wall structure can and does change during the course of the wall motion, indicating that the wall certainly cannot be regarded as a rigid object. Our results are consistent with the central ansatz of recent analytical theory \cite{Alouges02,DeSimone04}, in which emphasis is given to the stability of the geometry of the repeating unit in the XDW. Rather than suffering a significant distortion of the basic geometry, we observed the wall to simply generate new vortex/anti-vortex pairs as the cross-ties tails become shortened, thereby increasing the cross-tie density. Indeed, cross-tie densities almost an order of magnitude greater than those observed in the unconstrained case could be realised. Moreover, the increase in density observed as the wall approaches an edge results in expulsion of the wall, attained when the central wall portion touches the edge, being more difficult than might be anticipated. Similarly, difficulties are encountered when the field is relaxed and the energetically favoured state is for the wall to move back towards the middle of the element. However, the cross-tie density is now higher than optimal with the result that as the wall returns towards the centre it has to expel cross-ties. This turns out not to be an easy process, leading to a new source of hysteresis. Detailed study of this behaviour is underway and will be reported in a future publication.


\acknowledgments
The authors wish to thank Antonio DeSimone and Stefan M\"{u}ller for fruitful discussions and John Weaver and the James Watt Nanofabrication Centre for provision of facilities for electron lithography. This work has been supported by the “MULTIMAT” Marie Curie Research Training Network (MRTN-CT-2004-505226).


\end{document}